\documentclass[]{aastex631}

\begin{document}

\def\kms{$\rm km~s^{-1}$}
\def\cmcube{$\rm cm^{-3}$}
\def\kmss{$\rm km~s^{-1}$ }
\def\cmcubes{$\rm cm^{-3}$ }

\title{The Distance to the S147 Supernova Remnant}


\author[0000-0001-6017-2961]{C. S. Kochanek}
\affiliation{Department of Astronomy,
The Ohio State University \\
140 West 18th Avenue,
Columbus, OH 43210}
\affiliation{Center for Cosmology and Astroparticle Physics,
The Ohio State University \\
191 W. Woodruff Avenue,
Columbus, OH 43210}

\author{John C. Raymond}
\affiliation{Harvard-Smithsonian Center for Astrophysics \\
60 Garden Street,
Cambridge, MA 02138}

\author{Nelson Caldwell}
\affiliation{Harvard-Smithsonian Center for Astrophysics \\
60 Garden Street,
Cambridge, MA 02138}



\begin{abstract}
In the absence of a parallax distance to a pulsar or a surviving binary in a supernova remnant (SNR), distances to 
Galactic SNRs are generally very uncertain.
However, by combining Gaia data with wide field, multi-fiber echelle spectroscopy, it is now possible to obtain accurate distances to many SNRs with limited extinction by searching for the appearance of
high velocity Ca~II or Na~I absorption lines in hot stars as a function of distance.  We demonstrate this for 
the SNR S147 using the spectra of 259 luminous, blue stars. We
obtain a median distance of $1.37$~kpc ($1.30$ to $1.47$~kpc at
90\% confidence) that is consistent with the median parallax distance
to the pulsar of $1.46$~kpc ($1.12$ to $2.10$~kpc at 90\% 
confidence), but with significantly smaller uncertainties.  Our 
distance is also consistent with the distance to the candidate
unbound binary companion in this SNR, HD~37424. The presence of
high velocity absorption lines is correlated with the emission
line flux of the SNR but not with the radio flux.
\end{abstract}


\section{Introduction} \label{sec:intro}

Supernova remnants (SNR) dominate the dynamics of the ISM and the chemical evolution of the Galaxy.  They are physically interesting and important
probes of both the explosion and the nature of the progenitor
stars.  For example, estimates of ejecta masses and velocities constrain
the progenitor and explosions (e.g., \citealt{Temim2022}). They are
laboratories for shock physics (e.g., \citealt{Raymond2020a}, \citealt{Raymond2020b}).
They can be searched for bound or unbound
binaries after the explosion in both core collapse (e.g., \citealt{Kochanek2019}, \citealt{Kochanek2021})
and Type Ia remnants (e.g., \citealt{Kerzendorf2009}). Just as in extragalactic
SNRs (e.g., \citealt{Badenes2009}, \citealt{Jennings2014}), the age distribution of the surrounding stellar population can be
used to estimate the mass of the progenitor (\citealt{Kochanek2022}).

A major challenge in using SNRs for any of these
applications is determining their distances.  Almost
all Galactic SNR distances are significantly uncertain. The most common estimates
use HI velocities or pulsar dispersion measures, neither of which
are very precise.  For our eventual target, S147 (Simeis 147, G180.0$-$1.7, the Spaghetti nebula), 
the traditional distance estimates range from roughly 0.8 to 2~kpc (see the summary in 
\citealt{Dincel2015}), leading to factor of 2.5
uncertainties in sizes or velocities from proper motions and a factor of 6 in
luminosities or masses.  Only three Galactic SNRs have 
distances from pulsar parallaxes: Vela (\citealt{Dodson2003}), the Crab (\citealt{Lin2023}, both
VLBI and Gaia) and S147.  The pulsar
in S147, PSR~J0538$+$2817, has a parallax
of $0.72 \pm 0.12$~mas (\citealt{Ng2007}, \citealt{Chatterjee2009}) which
corresponds to a median distance of $1.46$~kpc ($1.12$-$2.10$~kpc at 90\% confidence).
There are also parallax distances to G039.7$-$02.0, G205.5$+$00.5 (Monoceros loop)
and G284.3$-$01.8
because they contain the interacting binaries SS~433 (see the review by
\citealt{Margon1984}), MWC~148 (\citealt{Hinton2009}) and
2MASS~J10185560$-$5856459 (\citealt{Corbet2011}), respectively, and the donor stars
have Gaia (\citealt{Gaia2016}, \citealt{Gaia2023}) parallaxes.
Curiously, the Gaia parallax for SS~433 of $\varpi=0.118\pm0.023$~mas
(geometric distance $6.4$-$8.5$~kpc, photogeometric distance of
$5.5$-$8.9$~kc, \citealt{BailerJones2021})
is at best marginally consistent with estimates from kinematic models
of its jets (e.g., $5.5\pm0.2$~kpc, \citealt{Blundell2004}).

\cite{Cha1999} demonstrated a new method for measuring SNR distances,
looking for the appearance
of high velocity (up to several 100~km/s)
Ca~II and Na~I absorption features from the SNR in the
spectra of hot stars (which lack
such features) superposed on the SNR as a function of distance.
They estimated a distance to
the Vela SNR of $250 \pm 30$~pc
which was later confirmed by the pulsar parallax distance of $286\pm16$~pc
(\citealt{Dodson2003}).  Interestingly, the high velocity absorption features are also
rapidly time variable in strength and wavelength (e.g., \citealt{Rao2020} most recently).  The only
other case where this approach has been used is for the Cygnus Loop
(\citealt{Fesen2018}, \citealt{Fesen2021}, \citealt{Ritchey2024}).

\cite{Cha1999} depended on Hipparcos (\citealt{Perryman1997}) parallaxes and so could
only do a very nearby SNR like Vela.  With Gaia, it is possible to apply
the method to tens of SNRs, as demonstrated by the Cygnus Loop result (\citealt{Fesen2018}, \citealt{Fesen2021}).
But both studies depended on observing stars one at a time, which leads
to either a very expensive program ($\sim 70$ stars for Vela) or a very
sparse sample ($6$ in each study for Cygnus).   Interestingly, only some background
stars show the high velocity absorption features.  It is not known if
the existence of absorption is
(anti)correlated with any observational property of the SNR (radio,
emission lines, X-rays, etc.).  Fortunately, there are now wide field,
multiobject, fiber fed, echelle spectrographs like Hectochelle on the
MMT (\citealt{Fabricant2005}, \citealt{Szentgyorgyi2011}), M2FS
on Magellan (\citealt{Mateo2012}), FLAMES on the VLT (\citealt{Pasquini2000}), and HERMES on the AAT
(\citealt{Sheinis2016}) which allow observations of tens to hundreds of 
stars simultaneously.

Here we demonstrate this approach using the MMT and Hectochelle to observe
stars in the direction of S147.  The one degree field of view of the
fiber positioner is well matched to the $3^\circ$ diameter (\citealt{Green2009})
of S147 and many other SNRs. Hectochelle can obtain 240 spectra at a resolution
of 34,000 ($\sim 8$~km/s) in 150\AA\ windows centered on Ca~II~$\lambda\lambda3933.663, 3968.47$ (H and K) or
Na~I~$\lambda\lambda5889.51,5895.924$ (D1, D2).  We chose S147 as a demonstration
case because the pulsar parallax (\citealt{Ng2007}, \citealt{Chatterjee2009}) provides a 
check of the method.  S147 is also of interest because it contains the one fairly convincing
example of a binary star, HD37424, unbound in a supernova explosion (\citealt{Dincel2015}, \citealt{Kochanek2021}).  We describe the selection of the stars and the observations
in \S2, and the observational results and analysis in \S3.  In \S4 we summarize the
results and discuss future applications.

\begin{figure}
    \centering
    \plottwo{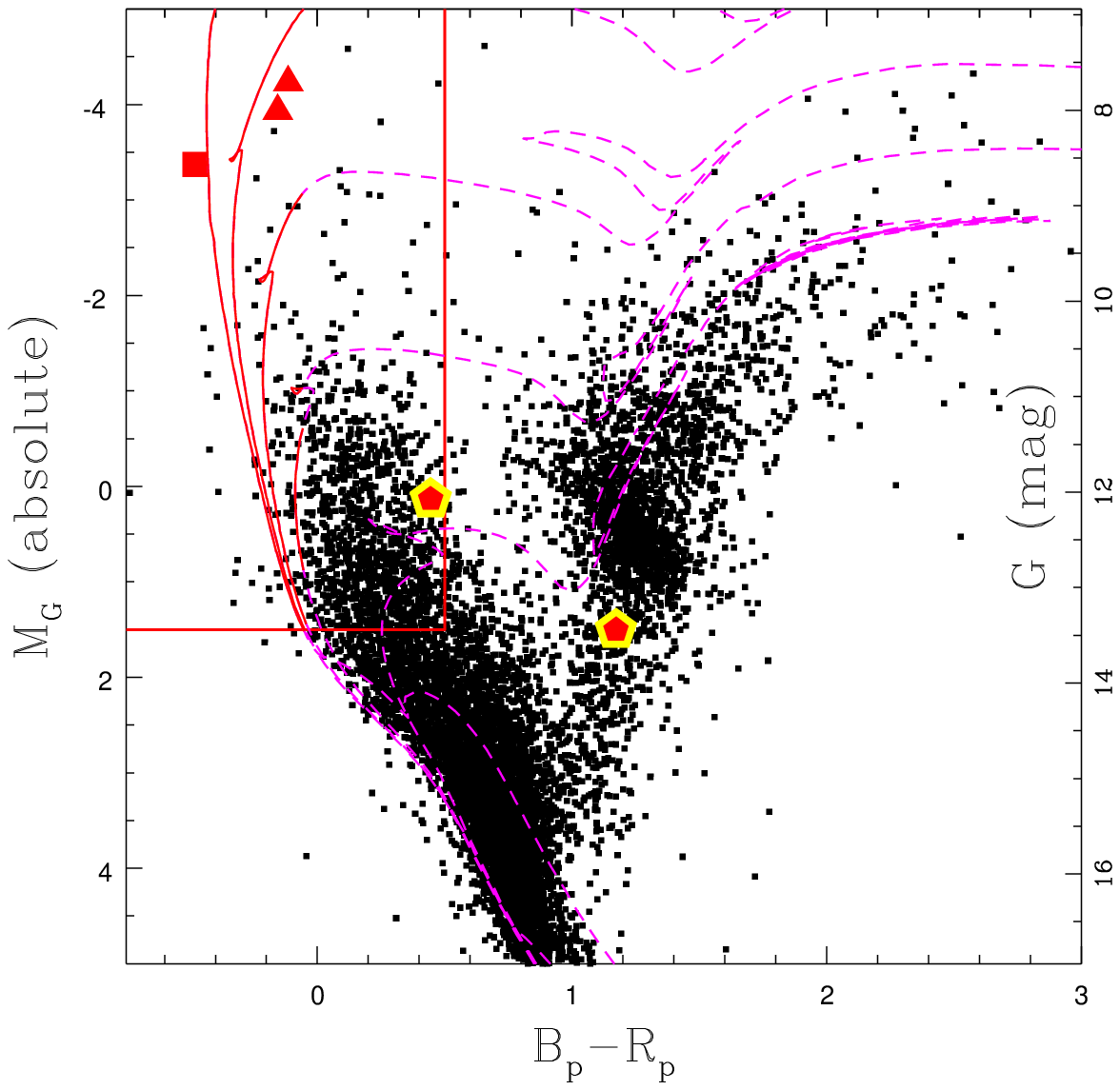}{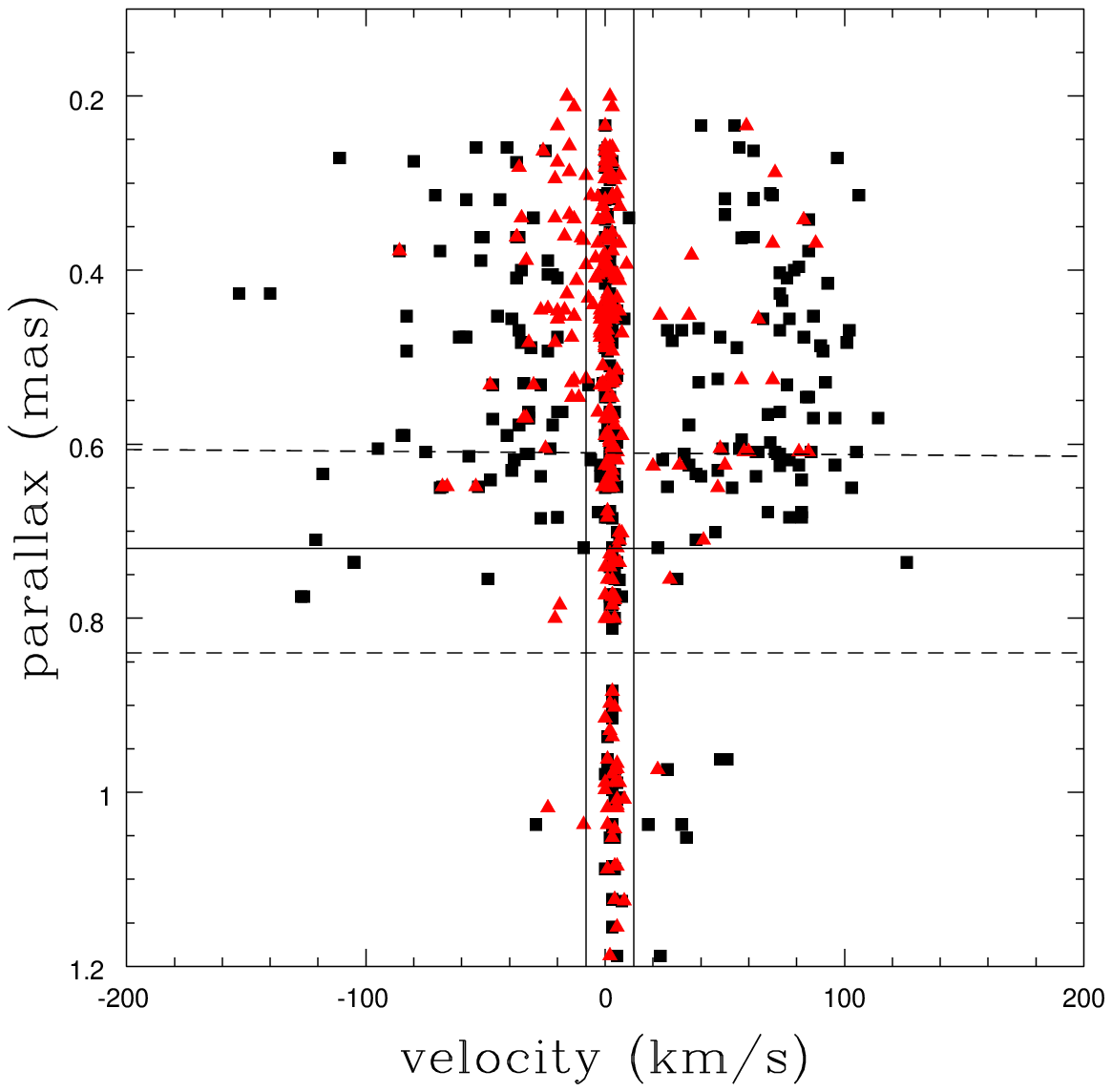}
    \caption{(Left) The distribution of the stars projected on S147 in absolute magnitude $M_G$ and
       $B_P-R_P$ color after correcting for extinction. The vertical scale
      on the right is the corresponding apparent magnitude at the parallax
      distance of the pulsar.  The curves are PARSEC isochrones from
      $10^7$ to $10^9$~years in steps of $0.5$~dex colored solid red (dashed magenta) for 
      temperatures $>10^4$~K ($<10^4$~K).  The red square is the candidate unbound
      binary companion, HD~37424, and the red triangles are two background stars,
      HD~36665 and HD~37318, with known high velocity absorption lines. Stars in the upper left corner
      outlined by the red box are our high priority targets. The yellow/red pentagons
      are the two stars whose spectra are shown in Figs.~\protect\ref{fig:abs58} (bluer star)
      and \protect\ref{fig:abs95} (redder star).
       }
    \label{fig:cmd}
    
    \caption{(Right) Heliocentric velocities of Ca~II (black squares) and Na~I (red triangles) absorption features
       as a function of parallax.  Note that the parallax axis is inverted so that more 
       distant stars are at the top.  The horizontal lines show the pulsar parallax and
       its uncertainties.  Absorption between the vertical lines at $2\pm 10$~km/s is viewed as being  due
       to the ISM for our standard model.
       }
    \label{fig:stars}
\end{figure}

\begin{figure}
    \centering
    \includegraphics[width=\linewidth]{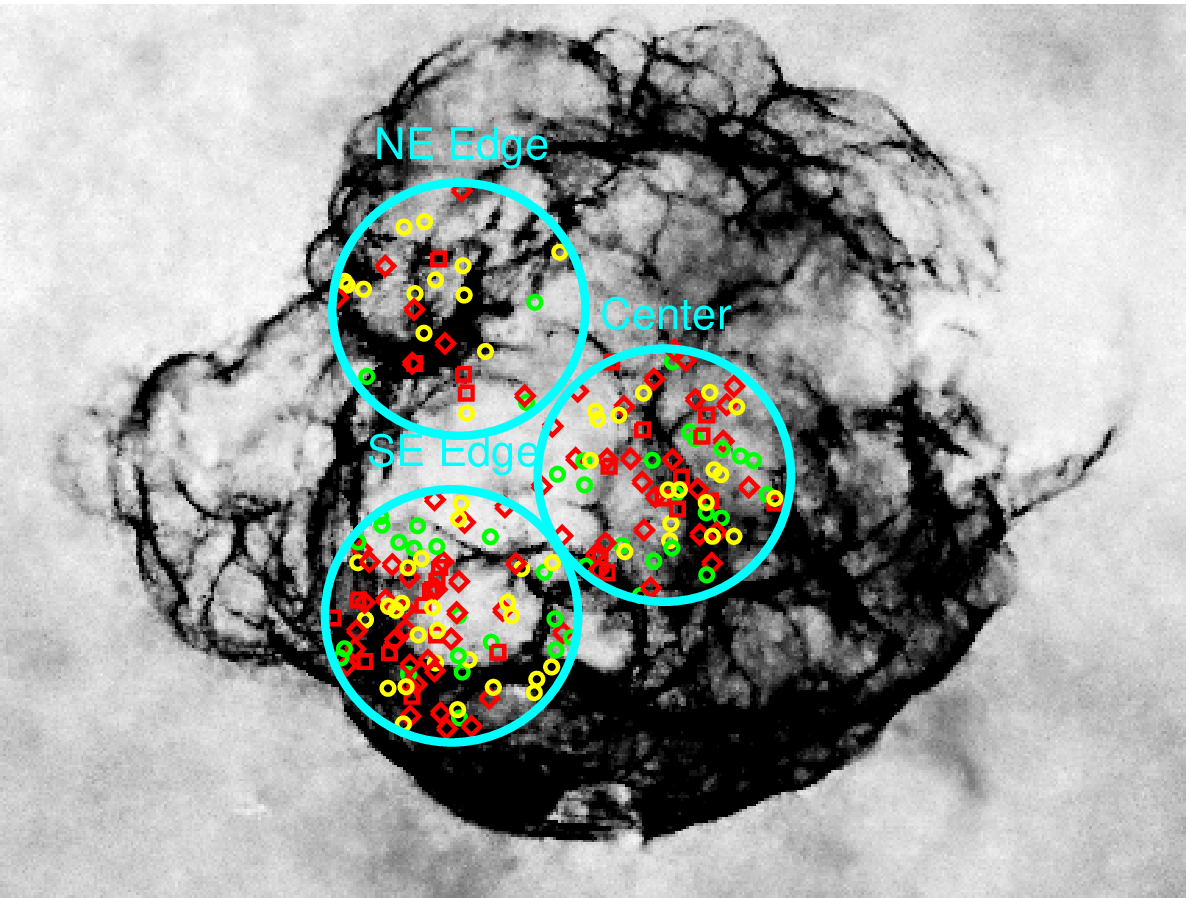}
    \caption{Background stars with both Na and Ca high velocity absorption features (red squares), only
       one of Na or Ca high velocity absorption features (red diamonds) or no high velocity
       absorption features (green circles) superposed on the 
       dual H$\alpha$/O[III] emission line image of S147.  The large cyan circles are the three $1^\circ$ diameter 
       Hectospec fields and the small yellow circles are foreground stars.
       }
    \label{fig:halpha}
\end{figure}

\begin{figure}
    \centering
    \includegraphics[width=\linewidth]{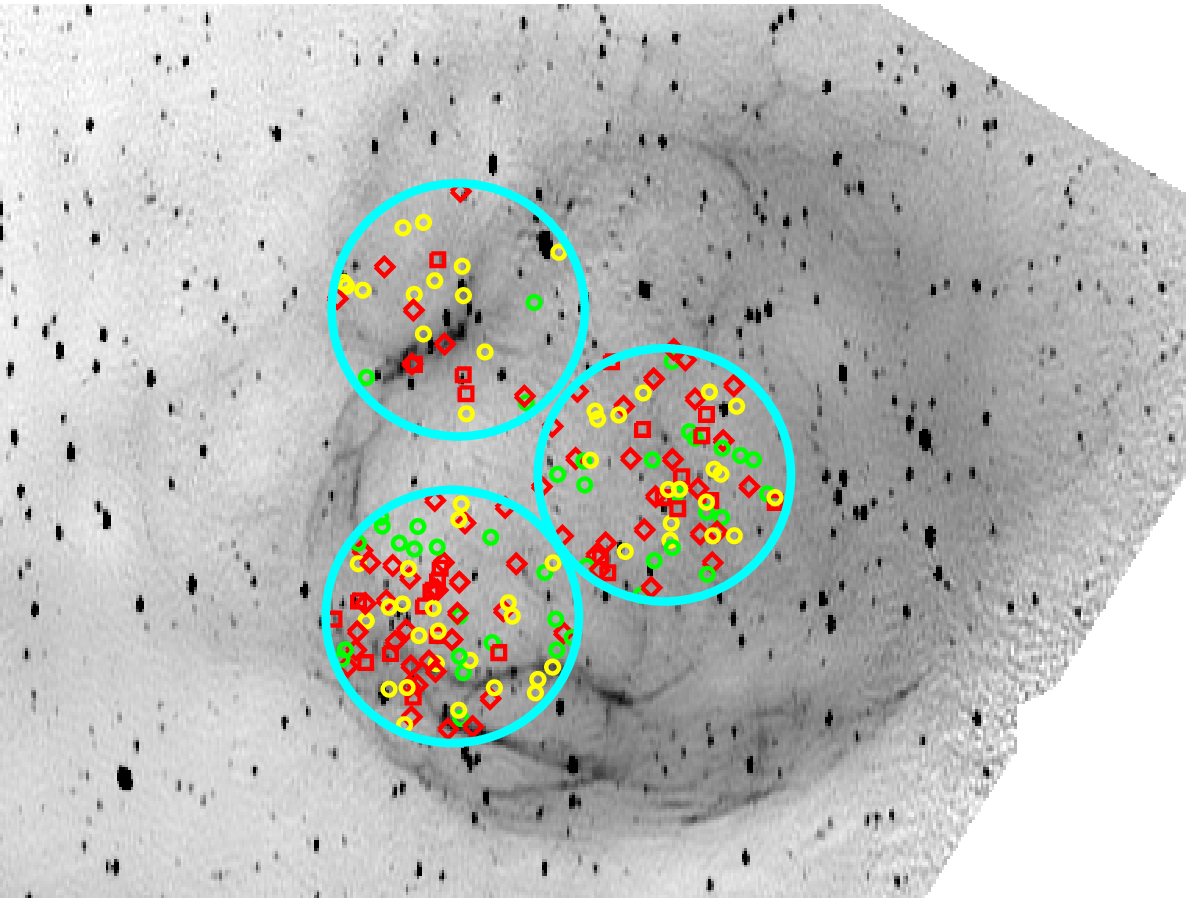}
    \caption{Background stars with both Na and Ca high velocity absorption features (red squares), only
       one of Na or Ca high velocity absorption features (red diamonds) or no high velocity
       absorption features (green circles) superposed on the 1.4~GHz Canadian Galactic Plane
        Survey map of S147.  The large cyan circles are the three $1^\circ$ diameter 
       Hectospec fields and the small yellow circles are foreground stars. Stars too close
        to the radio point sources are not shown. The scale
        of this image is the same as in Fig.~\protect\ref{fig:halpha}.
       }
    \label{fig:radio}
\end{figure}

\section{Sample and Observations} \label{sec:sample}

We selected stars from Gaia DR3 (\citealt{Gaia2016}, \citealt{Gaia2023}) within $1.7^\circ$
of the center of the SNR (RA $84.75^\circ$, Dec $27.83^\circ$, \citealt{Green2009}). They
were also selected to have $G<15$~mag and $\varpi>0.2$~mas.  We obtained estimates for 
the extinction towards each star using the three dimensional {\tt combined19 mwdust}
models (\citealt{Bovy2016}) which are based on the \cite{Drimmel2003},
\cite{Marshall2006} and \cite{Green2019} extinction distribution models.  Fig.~\ref{fig:cmd}
shows the distribution of the stars in extinction-corrected absolute magnitude and color
simply using the inverse parallax for the distance.  Superposed are Solar metallicity
PARSEC (\citealt{Bressan2012}, \citealt{Marigo2013}, \citealt{Pastorelli2020}) isochrones
with ages of $10^7$ to $10^9$ years in steps of $0.5$~dex colored red for temperatures
above $10^4$~K.  Also marked are the candidate unbound companion to the neutron star,
HD~37424, and two stars, HD~36665 and HD~37318, previously known to show high velocity
Ca/Na absorption lines (\citealt{Silk1973}, \citealt{Phillips1981}, \citealt{Phillips1983}, \citealt{Sallmen2004}). \cite{Dincel2015} were concerned that
these stars had distance estimates that could put them in the foreground of the SNR, 
but their Gaia parallaxes put them at or beyond the SNR distance (see \citealt{Kochanek2021}).

Stars hotter than roughly $10^4$~K (i.e., O and B stars) do not have Ca~II or Na~I
absorption (the coolest B stars can).  These are the optimal targets to search
for absorption lines from the interstellar medium (ISM) and the SNR.  We selected
as our primary targets stars with $M_G < 1.5$~mag and $B_P-R_P<0.5$~mag, as shown
in the red box in Fig.~\ref{fig:cmd}.  We also required total proper motions $\mu_{tot}<10$~mas/year.
Higher observational priorities were given to stars with
parallaxes $0.4 < \varpi < 2.0$~mas, as these should most strongly bracket the 
expected distance.  Objects outside the box were included at still lower 
priorities.  We divided the targets into Bright and Faint samples at
$G<11.5$~mag to help obtain similar counts over the large brightness 
range. The Bright sample was filled in with lower priority
$G>13$~mag stars just to avoid having empty fibers.

We used the three pointings 
shown in Figs.~\ref{fig:halpha} and \ref{fig:radio}.  One was centered on the 3\degr \ diameter SNR 
(``Center'') and the other two were centered contiguously to the southeast (``SE Edge'')
and northeast (``NE Edge'') of the center to capture the edges, where we might expect lower velocities 
but higher projected column densities.  Each was observed in two spectrograph configurations
to cover both the Ca~II and Na~I wavelength ranges.  We obtained a total of eleven 
observations.  We obtained all 4 basic observations (Bright/Faint, Ca~II/Na~I) for the
Center and SE Edge fields, plus one extra Bright, Na~I observation of the Center field.
We only obtained the Bright Ca~II/Na~I observations for the NE Edge field.
For the Faint observations, a star with $G \simeq 12.5$~mag had a signal-to-noise (S/N) ratio of 
$\simeq 60$ per pixel or $\simeq 300$ per resolution element.
For the shorter exposure Bright fields, the same S/N was achieved for stars with $G\simeq 11$~mag.
Table~\ref{tab:fields} summarizes the observations, where $N_{spectra}$ is the number of stellar
spectra obtained for each pointing.

The observations were processed using the standard IDL pipeline for Hectochelle data developed by the SAO 
Telescope Data Center (TDC\footnote{\url{https://lweb.cfa.harvard.edu/mmti/hectospec/hecto_pipe_report.pdf}}).  Briefly, the separate CCD images are combined with cosmic-ray elimination and the target traces are extracted to form 1D spectra. Wavelength calibration is supplied by ThAr exposures taken at the time of observation, and sky subtraction is performed using contemporaneous fiber spectra of blank regions within the 1 degree field. The wavelength stability of Hectochelle has been measured to be $0.78\pm 0.78$ \kms over 3 years, with an rms variation among fibers for a single pointing of just 0.17 \kms\ (P. Cargile, priv. comm.). In total , we obtained spectra of
552 stars, of which 259 had $M_G<1.5$~mag and $B_P-R_P<0.5$~mag (the red box region of Fig.~\ref{fig:cmd})
and 293 were either less luminous or redder.  Of the 259 luminous blue stars, 172 (194) were good spectra
showing clean Ca~II (Na~I) absorption lines, and 101 (60) of these stars had high velocity absorption lines (heliocentric velocities
outside $2\pm10$~km/s).  The velocities of these clean absorption lines are shown in Fig.~\ref{fig:stars} and listed in Table~\ref{tab:velocity}.

\begin{deluxetable}{lccrrrrr}
\tablecolumns{8}
\tablewidth{0pc}
\tablecaption{Field information \label{tab:fields}}
\tablehead{\colhead{Field/} 
&\colhead{RA}
&\colhead{Dec}
&\colhead{Line}
&\colhead{Time} 
&\colhead{$N_{spectra}$} 
&\colhead{G mag}
&\colhead{UT Date}\\
\colhead{Config}  &\multicolumn{2}{c}{(J2000)} & & \colhead{(s)} & &\colhead{range} & \\
 }
\startdata
Center/Faint  & 05:39:08.4 &   +27:48:06.2& Ca~II & 9900&201 & 11.5-15&2023-09-26    \\
Center/Faint  & " & " & Na~I & 2160&203 & 11.5-15&2023-09-25    \\
Center/Faint  &" & " & Na~I & 2160& 201 &11.5-15&2023-11-09    \\
Center/Bright  & " &  "  & Ca~II & 2160& 46 & 8.5-13&2023-11-04    \\
Center/Bright  & "&   " & Na~I & 900& 46& 8.5-13&2023-11-04    \\
SE~Edge/Faint  & 05:42:55.9 &   +27:14:48.5 & Ca~II & 9360& 203 & 11.5-15&2023-09-27    \\
SE~Edge/Faint  & " &  "& Na~I & 3600&203 & 11.5-15&2023-10-12    \\
SE~Edge/Bright  & "&   " & Na~I & 900& 56 & 8.5-13& 2023-11-04    \\
SE~Edge/Bright  & "&  " & Ca~II & 2160& 56 & 8.5-13&2023-11-04    \\
NE~Edge/Bright  & 05:42:49.1& +28:27:27.9  & Na~I & 1080& 46 & 8.5-13& 2023-11-08    \\
NE~Edge/Bright  & "&  " & Ca~II & 2160& 46 & 8.5-13&2023-11-24    \\
\hline
\enddata

\end{deluxetable}

\begin{figure}
    \centering
    \plottwo{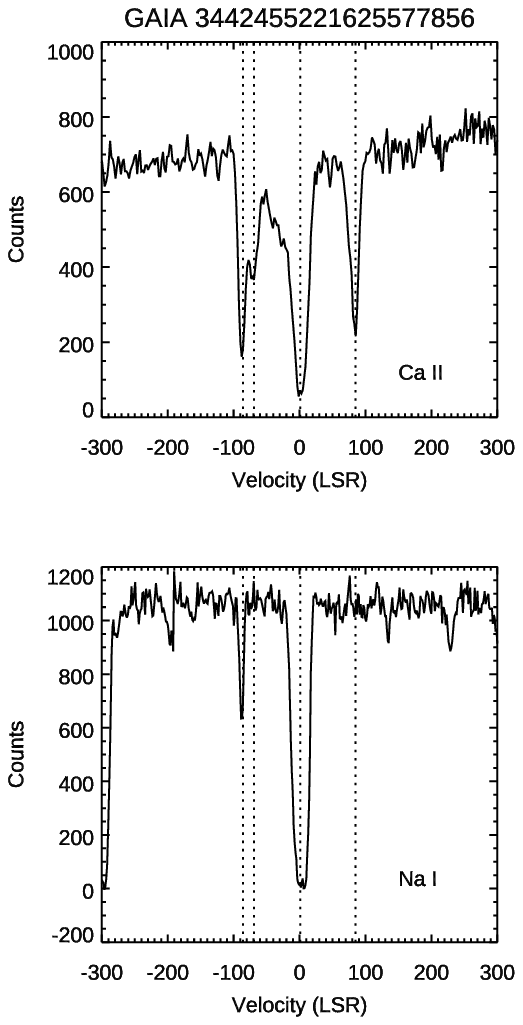}{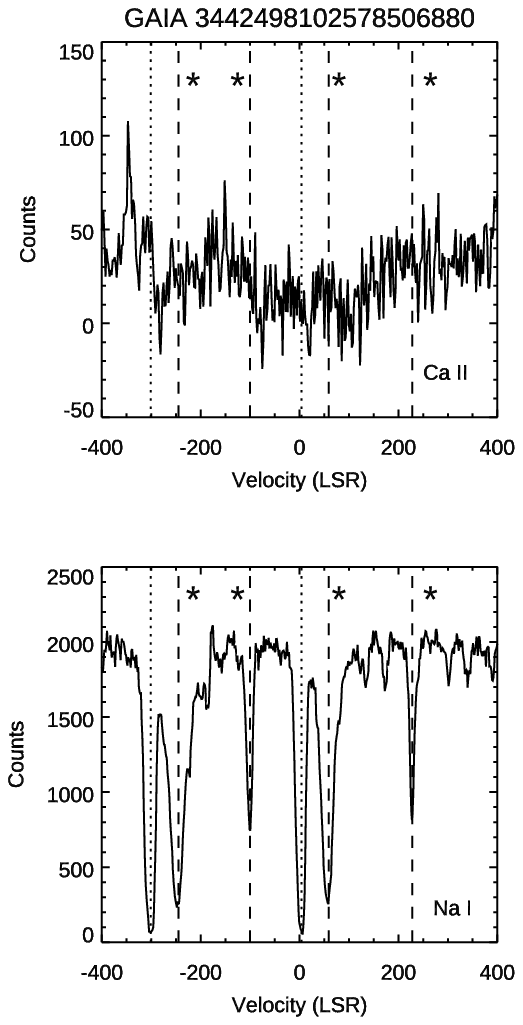}
    \caption{(Left) The Ca~II and Na~I spectra of the luminous, blue star near the lower, right corner
    of the selection box in Fig.~\protect\ref{fig:cmd}.  This star has strong ISM absorption near
    zero velocity for both species. There are three additional strong Ca~II absorption features at 
    $-86$, $-69$, and $+85$~km/s, while only the feature at $-86$~km/s is seen in the Na~I spectrum. The strong absorption line at $-300$~km/s is the other component of the
    interstellar Na~I doublet at $0$~km/s.
       }
    \label{fig:abs58}
        \caption{(Right) The Ca~II and Na~I spectra of the cool star just below the red clump in
        Fig~\protect\ref{fig:cmd}.  There are few counts in the Ca II spectrum because the star
        is a red giant. The high-signal-to-noise ratio Na~I spectrum has strong interstellar
        absorption near $0$~km/s, but also strong stellar absorption lines at $+59$~km/s of
        multiple species including Na~I, as labeled and marked by the dashed lines and asterisks.
         }
    \label{fig:abs95}
\end{figure}

The spectra were individually examined to identify the velocities of Ca~II and Na~I absorption features, which
were flagged as being clean or not.  Since equivalent widths, particularly with the complications of
overlapping, weaker stellar and interstellar features, are unimportant for our present objectives,
we simply measured line centroids. The Ca~II absorption lines were generally stronger than the corresponding
Na~I absorption lines.  The biggest distinction is simply between the luminous, blue, primary targets and 
the fainter or redder low priority and filler targets.  Fig.~\ref{fig:abs58} shows the spectra of a primary
target near the lower (reddest), right (faintest) corner of the red selection box in Fig.~\ref{fig:cmd}. 
The star has strong interstellar absorption near zero velocity and then strong Ca~II lines near
$-86$, $-69$ and $+85$~km/s.  Only the $-86$~km/s line appears in the corresponding Na~I spectrum.
Fig.~\ref{fig:abs95} shows the spectra of a red giant star a little below the red clump in Fig~\ref{fig:cmd}.
The low blue flux of the red giant leads to a very noisy Ca~II spectrum. The Na~I spectrum has a high signal-to-noise
ratio, but in addition to the interstellar Na~I absorption, there is stellar Na~I absorption plus
additional stellar atmospheric absorption lines of other species.  For these stars, high velocity
interstellar absorption can only be identified after modelling the spectrum to identify and remove
these stellar features.  For this paper we only use the luminous blue stars with clean 
absorption features.

\begin{figure}
    \centering
    \plottwo{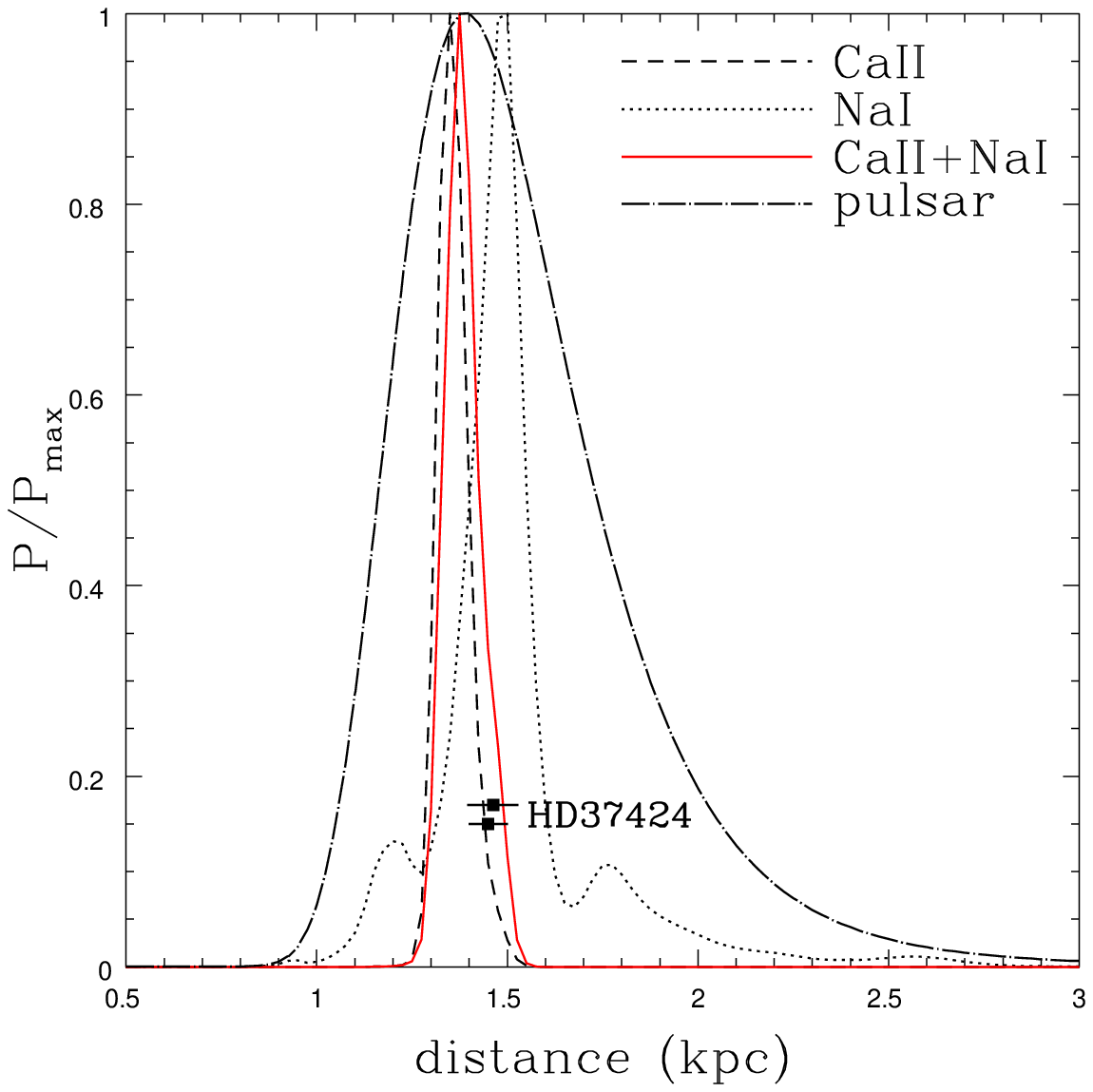}{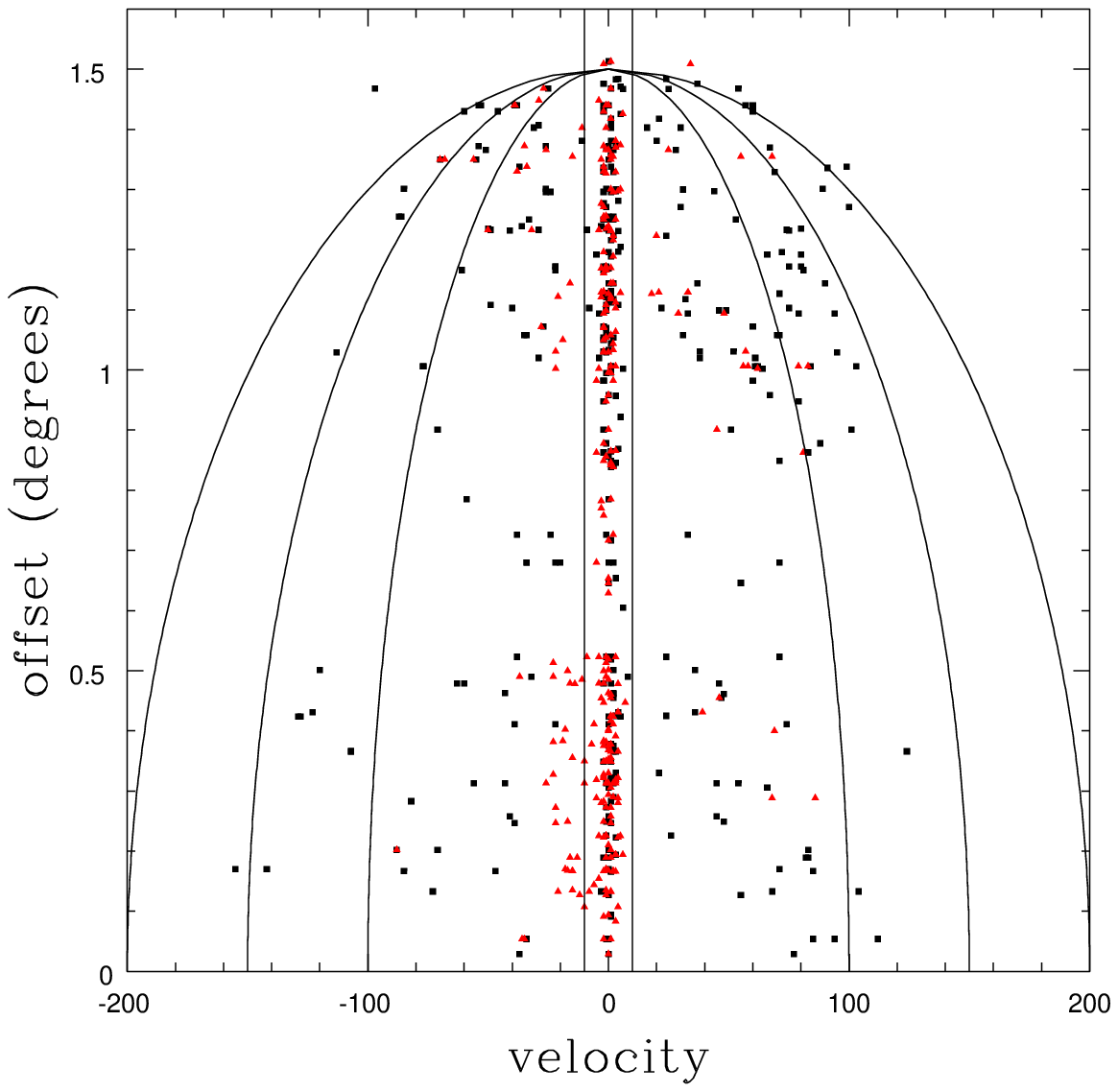}
    \caption{(Left) Normalized $P/P_{max}$ probability distributions for the
    distance to S147 using only the Ca~II (CaII, black dashed), only the Na~I (NaI, black dotted) or both (CaII$+$NaI, red solid) absorption lines for an ISM velocity window of $2\pm 10$~km/s.
     The black solid curve is the distance probability
    distribution implied by the pulsar parallax. The points show 
    the \protect\cite{BailerJones2021} photometric (upper) and
    photogeometric (lower) distances to the candidate unbound
    binary companion HD~37424. 
       }
    \label{fig:dist}
        \caption{(Right) Absorption velocities relative to the LSR as a function of offset
        from the center of the SNR.  The curves show the maximum expected velocities (Eqn.~\ref{eqn:expand})
        for a radius of $R_{SNR}=1.5^\circ$  and velocities at the edge of 
        $v_0=100$, $150$ and $200$~km/s. The
        median 2~km/s LSR velocity has been subtracted.
       }
    \label{fig:kinematics}
\end{figure}

\begin{figure}
    \centering
    \plottwo{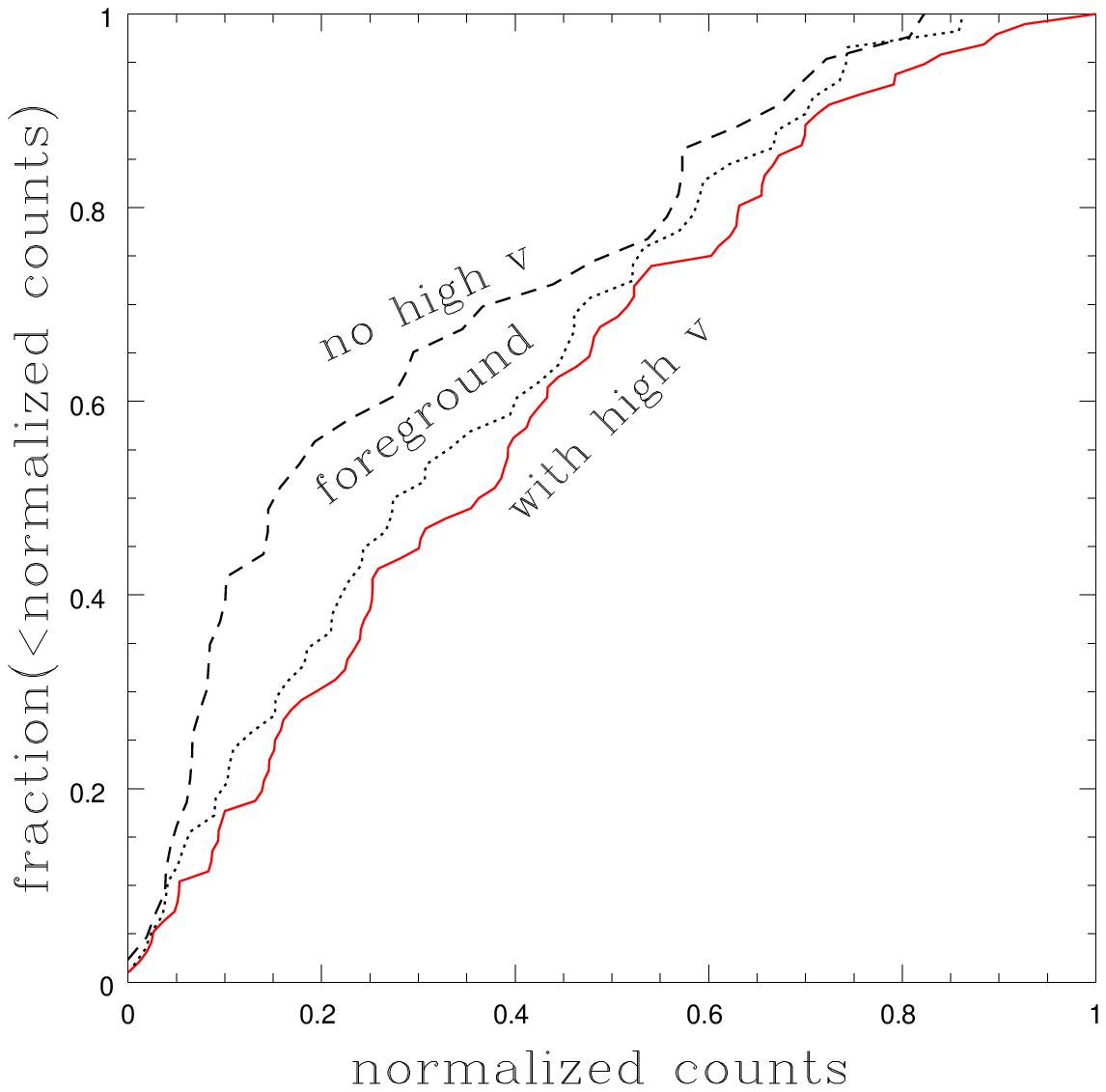}{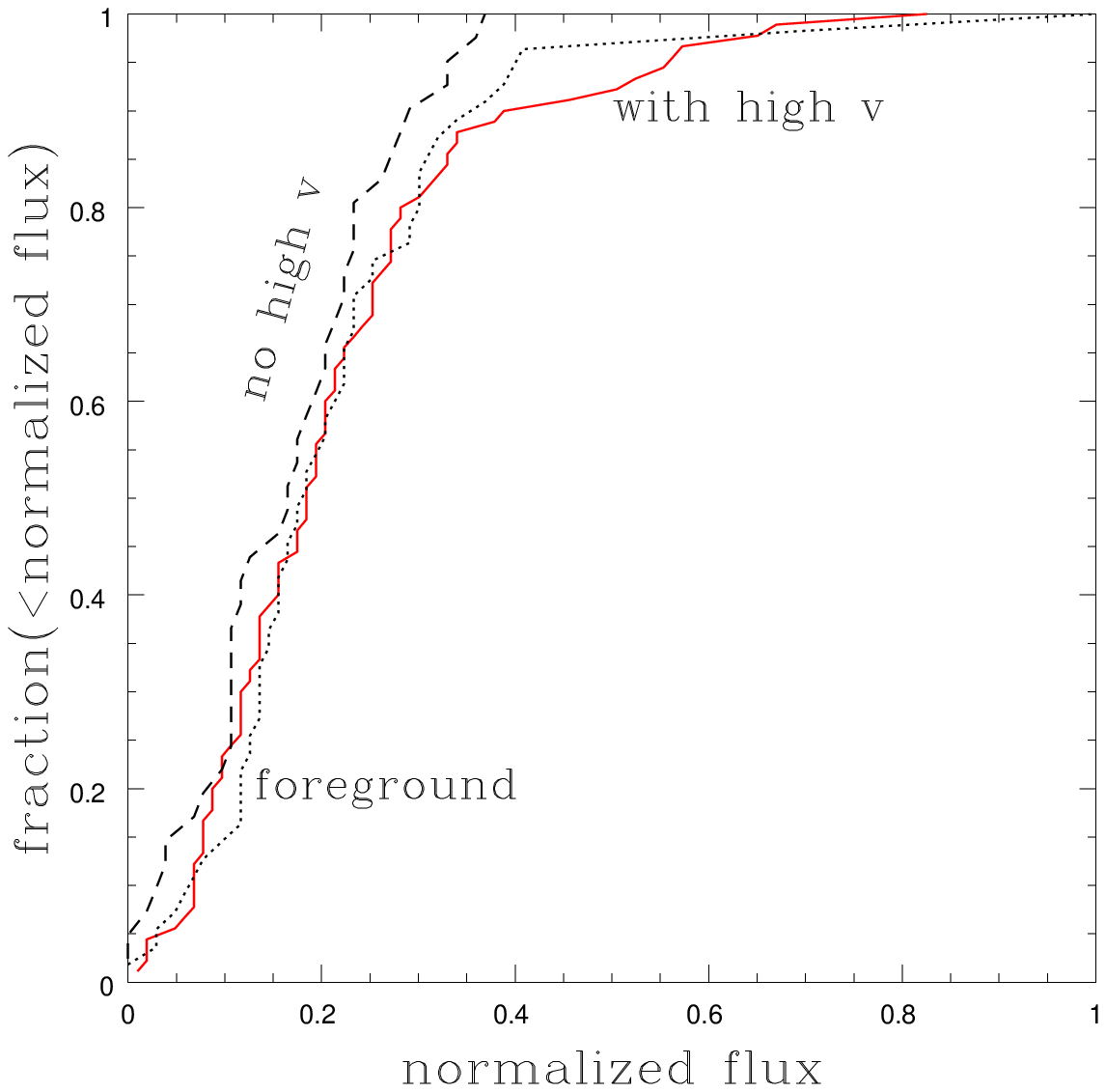}
    \caption{(Left) Integral distributions of the background stars with high velocity absorption (solid red) and without
       high velocity absorption (dashed black) in normalized counts from the emission line image in
       Fig.~\protect\ref{fig:halpha}.  The dotted curve shows the distribution of the foreground stars.
        The Kolmogorov-Smirnov
       test probability that
       the two background distributions are the same is $1.5\%.$
       }
    \label{fig:halphaks}
        \caption{(Right) Integral distributions of the background stars with high velocity absorption (solid red) and without
       high velocity absorption (dashed black) in normalized flux from the 1.4~GHz image in
       Fig.~\protect\ref{fig:radio}.  The dotted curve shows the distribution of the foreground stars.
        The Kolmogorov-Smirnov
       test probability that
       the two background distributions are the same is 53\%.
       }
    \label{fig:radioks}
\end{figure}

\section{Results} \label{sec:results}

Fig.~\ref{fig:stars} shows the velocities of the clean absorption components in the luminous, blue stars
as a function of parallax.  There is a tight band at low velocities due to the ISM.
Because S147 lies almost directly towards the Galactic anti-center, there 
is no systematic distance dependence to the ISM velocity relative to the Sun.
If we roughly exclude the higher velocity absorption lines, the median of the ISM
velocity band is $2$~km/s. For our standard models we will assign
absorption lines within $2\pm 10$~km/s to the ISM, although we report
the consequences of different choices for the velocity width used to
define the interstellar absorption
There are some higher velocity features for stars closer than the pulsar
 of unknown origin, but there is a clear change in the numbers of
 high velocity absorption components at the parallax of the pulsar, with velocities approaching
$\sim 150$~km/s.   
\cite{Kirshner1979} and \cite{Ren2018} found a similar velocity range
range for the emission lines from the SNR.  

The goal is to determine the distance $d_0$ (in kpc) to the SNR, but we fit the
measured parallaxes of the stars using $\varpi_0=1/d_0$. The property of the SNR is that it produces
a change in the statistics of high velocity absorption lines, so we 
model the probability of a star with distance $d>d_0$ ($d<d_0$) 
having a high velocity absorption line as $p_{far}$ ($p_{near}$). For a given
distance, there are $n_h$ ($m_h$) $d>d_0$ ($d<d_0$)
stars with a high velocity absorption and $n_l$ ($m_l$)
without a high velocity feature. The probability of the observed distribution is then
\begin{equation}
 P(D| d_0, p_{far}, p_{near}) =
     p_{far}^{n_h} \left( 1 - p_{far} \right)^{n_l}
     p_{near}^{m_h} \left( 1 - p_{near} \right)^{m_l}.
     \label{eqn:prob1}
\end{equation}
Arguably, we should include the statistical weights so that
\begin{equation}
 P(D| d_0, p_{far}, p_{near}) =
     { (n_h+n_l)!(m_h+m_l)! \over n_h! n_l! m_h! m_l! }
     p_{far}^{n_h} \left( 1 - p_{far} \right)^{n_l}
     p_{near}^{m_h} \left( 1 - p_{near} \right)^{m_l},
     \label{eqn:prob2}
\end{equation}
and we comment on the consequences of doing so below.
We assume uniform priors of $0 \leq d_0 \leq 10$~kpc, $0 \leq p_{far} \leq 1$ and
$0 \leq  p_{near} \leq 1$ and use Bayesian methods to compute the final likelihood
distributions.  We include one additional elaboration to incorporate parallax
uncertainties.   For a star with parallax $\varpi_i \pm \sigma_i $ and a high velocity absorption line, we make its contribution
to (for example) $n_h$ be
\begin{equation}
   \delta n_h = { 1 \over \sqrt{2\pi} \sigma_i} \int_0^{\varpi_0} d\varpi 
        \exp\left( - { 1 \over 2 }\left( { \varpi - \varpi_i \over \sigma_i }\right)^2 \right)
\end{equation}
and its contribution to $m_h$ would then be $1-\delta n_h$.  This has the particular effect
of down weighting the importance of stars with large parallax errors.

Fig.~\ref{fig:dist} shows the resulting estimates for the distance
after marginalizing over the probabilities $p_{far}$ and $p_{near}$ of having a high 
velocity absorption line.  The results using only the Ca~II,
only the Na~I or both sets of absorption features give 
consistent results. The Na~I results are significantly worse
simply because fewer high velocity Na~I features were detected.
The median distance estimates for Ca-only, Na-only and
both are $1.34$, $1.47$, and $1.37$~kpc
with 90\% confidence regions of $1.28$-$1.43$, $1.18$-$1.94$
and $1.30$-$1.47$~kpc, respectively.  If we increase the 
width of the velocity window defining ISM absorption to
$2\pm15$~km/s or $2\pm20$~km/s, the median distances are both $1.36$~kpc
with 90\% confidence ranges of
$1.29$-$1.45$~kpc and $1.28$-$1.46$~kpc, respectively,
so the distance estimate is robust to changes in the definition
of low and high velocities.  Similarly, if we use Eqn.~\ref{eqn:prob2}
instead of Eqn.~\ref{eqn:prob1}, the median distances combining
both lines for the
three velocity ranges are $1.37$, $1.35$ and $1.36$~kpc,
respectively, with 90\% confidence ranges of $1.31$-$1.43$,
$1.30$-$1.41$ and $1.30$-$1.42$~kpc, so this choices matters
only at the level of $\sim 0.01$~kpc. For the parallax
of the pulsar ($\varpi=0.72\pm0.12$~mas),  the median distance
is 1.46~kpc with a 90\% range of $1.12$-$2.10$~kpc  

Fig.~\ref{fig:kinematics} shows the absorption line velocities relative
to the LSR as a function of each star's offset $R$ from the center of the 
SNR.  If the absorbing material is in spherical, homologous expansion, then 
the velocities should be confined within the curve
\begin{equation}
    v = v_0 \left( 1 - {R^2 \over R_{SNR}^2 }\right)^{1/2}
      \label{eqn:expand}
\end{equation}
where $v_0$ is the expansion velocity at $R_{SNR}$.  Such curves are shown
in Fig.~\ref{fig:kinematics} for $v_0=100$, $150$ and $200$~km/s
and $R_{SNR}=1.5^\circ$ ($34$~pc given our median distance estimate).  If we adopt $v_0=150$~km/s and assume
Sedov-Taylor expansion, then the age of the remnant is $94.6\pm 13.4$
thousand years with $\log(E_{51}/n_1)=0.36\pm 0.16$, where $E_{51}$
is the explosion energy in units of $10^{51}$~erg,
$n_0$ is the ISM density in units of cm$^{-3}$ and we have
assumed 10\% errors on the radius and velocity. These are similar
to the parameters proposed from the eROSITA X-ray observations
of S146 (\citealt{Michailidis2024}).  This age is moderately
longer than the kinematic age estimated using the pulsar distance from the
SNR center and its proper motion and significantly shorter than the pulsar spin down
age ($33_{-9}^{+17}$ and 618 thousand years, respectively, \citealt{Kramer2003}).  The
low ionization atomic Ca~II and Na~I states are expected only in the colder, outermost
shell of the SNR and the complex emission line morphology in Fig.~\ref{fig:halpha} 
suggests that the observed velocities may be subject to projection effects
where we see a higher absorption column density along lines of sight perpendicular to 
shock fronts.  This will tend to bias the observed absorption velocities to be lower
than for a purely spherical expansion.

Finally, we looked for correlations between the presence of high velocity
absorption lines and the SNR radio and emission line fluxes. Unfortunately, there is no
sufficiently deep, publicly available X-ray image of S147 (the eROSITA image of
\citealt{Michailidis2024} is probably suitable but it is not public).  For 
H$\alpha$ we first tried the map from \cite{Finkbeiner2003} which
has the limitation that its resolution is only 6'.  We were 
serendipitiously contacted by an astrophotographer (C. Mauche)
who had just generated an H$\alpha$ and O~[III] image of S147
using an ZWO~ASI29MC Pro camera and an HUTECH/IDAS NBZ UHS dual narrow-band filter.
The stars are removed by an 
artificial intelligence application in {\tt StarNetV2} and the image was binned to 18\arcsec .
We used the 1\arcmin\ resolution 1.4~GHz map from the Canadian Galactic Plane Survey
(CGPS, \citealt{Taylor2003}).  Figs.~\ref{fig:halpha} and
\ref{fig:radio} 
 show the dual emission line and radio images with the three
Hectospec fields and the target stars coded by whether they are
foreground, background stars with high velocity absorption or
background stars without high velocity absorption.

Figs.~\ref{fig:halphaks} and \ref{fig:radioks} show the 
normalized emission line and radio flux distributions for
each class of stars.  The distributions in emission line flux for the background
stars with and without high velocity absorption are strikingly
different  - stars with high velocity absorption are correlated
with brighter line emission.  The distribution of the
foreground stars then lies between the two background star
distributions. The Kolmogorov-Smirnov test likelihood that
the two background distributions are drawn from the same 
distribution is 1.5\%.  The emission line image
in Fig.~\ref{fig:halpha} was obtained as an astrophotography
project and so is not a clean H$\alpha$ or O~[III] image.  However,
if we use the well-defined, but much lower resolution \cite{Finkbeiner2003}
H$\alpha$ image, we find the same result.  The distributions are similar to
Fig.~\ref{fig:halphaks} but the Kolmogorov-Smirnov significance
of the differences is 21\%, almost certainly due to the much
lower resolution.  For the normalized radio flux distribution
in Fig.~\ref{fig:radioks}, we removed stars 
too close to the numerous point sources seen in Fig.~\ref{fig:radio}. 
While the resulting flux distributions in Fig.~\ref{fig:radioks} show the same basic ordering
as the emission line flux distributions in Fig.~\ref{fig:halphaks}, the differences are far smaller and
the Kolmogorov-Smirnov significance is only 53\%.

\section{Discussion}  \label{sec:discuss}

We believe that the idea from \cite{Cha1999} for the Vela SNR that the appearance of high velocity Ca~II and Na~I in stars as a function distance can be used to determine
the distance to SNRs is now well-validated by this work.  We obtain a distance to S147 of
$1.37$~kpc ($1.30$-$1.47$ at 90\% confidence), consistent with the less
precise pulsar parallax and the parallax of the candidate unbound binary
companion to the SN progenitor.  As seen in Fig.~\ref{fig:halpha}, we obtained
only one partial and two complete Hectospec pointings covering only $\sim 20\%$ of the remnant, so 
it would be easy to improve the statistics (Fig.~\ref{fig:stars} and \ref{fig:dist}),
to improve the kinematics of the SNR (Fig.~\ref{fig:kinematics}), or
to better study correlations between the absorption and other properties of the
SNR (Figs.~\ref{fig:halphaks} and \ref{fig:radioks}).

One of our goals was to determine which absorption lines were better
for this method, and it appears to clearly be the Ca~II lines. Of the
259 luminous hot stars observed ($M_G<1.5$~mag, $B_P-R_P<0.4$~mag), 101 showed high velocity Ca~II absorption lines and only 60 showed high velocity Na~I absorption.   The Ca~II absorption velocities are also systematically higher, with a median velocity after 
excluding the ISM
lines of $58$~km/s relative to the LSR, versus $25$~km/s for Na~I. This is
a result of needing higher total column densities to detect Na~I than Ca~II, because Na I is easily photoionized by near UV photons below 2400 \AA, while Ca II requires photons shortward of 1045 \AA . Ca~II and Na~I originate in radiative shock waves in the outer layers of the SNR \citep{Cox1972b}.  Projection effects lead to lower column densities and higher velocities normal to the shock front and the
reverse parallel to the shock front.  This means that an absorption feature requiring
a higher column density will be biased towards lower average velocities.  
While Ca~II shows more and stronger absorption, it has the 
disadvantage of a bluer wavelength (3934/3968\AA\ versus
5890/5896\AA\ for Na~I), where extinction losses are higher and 
spectrographs/detectors tend to be less efficient. 

We focused on luminous, hot stars because they lack stellar
Na~I and Ca~II absorption (and narrow stellar absorption features generally)
due to their temperatures.  This greatly simplifies identifying
absorption lines from the interstellar medium and the SNR.
It is possible to apply this method to Na~I absorption in
the more numerous, luminous red stars, but it is complicated by the stellar 
absorption features of Na~I and many other species, as
seen in Fig.~\ref{fig:abs95}. Essentially,
the stellar spectrum has to be modelled to confidently
distinguish interstellar and stellar absorption features.
This is almost certainly feasible, but beyond our present scope.
Being able to apply this method to luminous red giants will be
increasingly important for SNRs with higher extinctions.

In old SNRs such as S147, we expect strong correlations between the Ca~II and Na~I column 
and the H$\alpha$ and radio fluxes.  The radiative shock waves responsible for the
Ca~II and Na~I also drive the H$\alpha$ emission \citep{Cox1972a}, and the brightest radio flux in old SNRs originates in radiative shocks where the cooling, post-shock gas compresses both the ambient cosmic ray electrons and the magnetic field \citep{vanderLaan1962, Raymond2020a, Tutone2021}.  
We correlated the background stars with and without high velocity 
absorption lines with emission line and radio images of S147.  We
found a significant correlation of high velocity absorption with
the flux in an H$\alpha$/O[III] astrophotography image of S147 and,  more
weakly, with the far lower resolution H$\alpha$ image from \cite{Finkbeiner2003}.   
This suggests that using emission line maps to prioritize targets is likely more important
than our center versus edge experiment. We
found no significant correlation between the background stars with
high velocity absorption lines and the Canadian Galactic Plane
Survey (\citealt{Taylor2003}) 1.4~GHz radio map of S147.  The insignificant
correlation with the radio emission may be driven by a combination of the
modest (1\arcmin) resolution and limited dynamic range of the CGPS radio map
for the emission from the SNR.
Unfortunately, no sufficiently deep, publicly available X-ray image of S147 was
available to check for correlations.   

Based on these results, we initiated a program to obtain data for up to 
sixteen additional Northern and Southern SNRs using MMT/Hectospec and Magellan/M2FS 
in the first trimester of 2024.  As emphasized in the introduction, modern
multi-fiber, echelle spectrographs allow these observations to be carried
out very efficiently. The selection of target stars for this trimester
is not making use of the correlation between the existence of high 
velocity absorption and emission line flux we find here, but will certainly 
serve to confirm the correlation for use in future observations. We are
also investigating modeling the spectra of the red giants, which are
currently just used to fill otherwise unused 
fibers, to also use them in the search for absorption features from the SNRs.

\begin{acknowledgments}
We thank Christopher Mauche for sharing his images of S147 with us. We also
had helpful discussions regarding sodium absorption with Jesse Han.
CSK is supported by NSF grants AST-1908570 and AST-2307385. Observations reported here were obtained at the MMT Observatory, a joint facility of the Smithsonian Institution and the University of Arizona. Data was reduced by the SAO TDC; we thank Sean Moran and Jaehyon Rhee for their help in that work.
\end{acknowledgments}

\vspace{5mm}
\facilities{MMT}

\bibliography{ms}{}
\bibliographystyle{aasjournal}

\begin{deluxetable}{llllll}
\tablecaption{Absorption Velocities of the Hot, Blue Stars\label{tab:velocity}}
\tablehead{
\colhead{Gaia DR3 ID}
&\colhead{Ca~II}
&\colhead{Na~I} 
&\colhead{Gaia DR3 ID}
&\colhead{Ca~II}
&\colhead{Na~I} \\
  &\colhead{(km/s)} &\colhead{(km/s)} 
& &\colhead{(km/s)} &\colhead{(km/s)} \\
}
\startdata
\hline
3441702193599182336 & $-118$,$4$,$38$&$2$ &3441703598051819776 & $3$&$2$\\
3441706518633166080 & &$-21$,$0$ &3441723496636807040 & &\\
3441698860704616064 & & &3442475974907331200 & $3$&$0$\\
3442474561861744000 & &$-13$,$0$ &3442449552266565888 & $3$&$5$\\
3442476112346300800 & $3$&$2$ &3442452197968495616 & $-105$,$5$&$3$\\
3442455011170611712 & $5$&$5$ &3442479337865371392 & $-32$,$1$,$87$&$-34$,$0$\\
3442469751498370560 & $-127$,$2$&$3$ &3442472779451400448 & $3$&$0$\\
3442491947889351296 & $3$&$4$ &3442491814744678528 & &$5$\\
3442546824687824256 & $0$&$3$ &3442506348915651328 & $5$,$23$&$2$\\
3442599669964938368 & $-58$,$0$&$-14$,$-2$ &3442503527120074624 & &$-24$,$1$\\
\hline
\enddata
\tablecomments{The entries are the velocities of the Ca~II and Na~I absorption
 lines separated by commas. 
 The complete table is available in the electronic version
 of the paper.}
\end{deluxetable}



\end{document}